\documentclass{article}
\usepackage{graphicx} 
\usepackage{amsmath}
\usepackage{amssymb}
\usepackage{natbib}
\usepackage{amsthm}
\usepackage{mathrsfs}
\usepackage{epigraph}
\usepackage{url}
\usepackage{booktabs}

\theoremstyle{definition}
\newtheorem{dfn}{Definition}[section]

\newtheorem{prin}[dfn]{Principle}

\title{Toward a Metaphysics of Learning Analytics: Ontological Positioning of Data, Inference, and Normativity}
\author{
    Kensuke Takii\thanks{Center for Educational Research and Development in AI and Data Science, Naruto University of Education, Naruto, Japan} \footnote{kensuke.takii96@gmail.com}
}
\date{}

\begin{document}

\maketitle

\begin{abstract}
    The Learning Analytics (LA) community has undergone rapid development over the 15 years since the first LAK conference was held. However, while epistemological and ethical debates regarding the philosophical foundations of LA have been vigorous, metaphysical discussions have been sparse, signifying a lack of effort to derive the identity of LA from its internal principles. In this paper, we attempt to establish a metaphysics of LA by addressing the ontological question of ``What is LA?'' We do so by tracing back to LA's own definitions and principles to derive an answer from within LA itself. Specifically, we address what kind of existence the data LA operates on constitutes, identify eight agents including learners as ontological prerequisites, and clarify, via the is/ought problem, that LA does not derive norms from data. In particular, this system reveals that a class of LA practices, here termed \textit{norm-embedded LA}, conflates LA's purpose with its operations, creating an ontological tension with the first principle. We also discuss connections with related fields and the limitations of this system. The metaphysics outlined here is not imposed from outside LA, but surfaces what LA itself has always implicitly presupposed.

\end{abstract}


\section{Introduction} 

\addcontentsline{toc}{section}{Introduction} 

\begin{quote}
    Learning analytics is the measurement, collection, analysis and reporting of data about learners and their contexts, for purposes of understanding and optimising learning and the environments in which it occurs.
\end{quote}
This is the definition of Learning Analytics (LA) first established in 2011 \citep{LAK11}. Riding the wave of the advent and rise of the big data era \citep{anderson2008end}, LA---which seeks to understand and improve learning using ``data about learners and their contexts''---has undergone rapid development \citep{SoLARWhatIsLA2025} from its relatively small beginnings \citep{siemens2013learning}. The field has since established a substantial research infrastructure, including dedicated conferences, journals, and handbooks. In practice, information systems and frameworks for LA are now in place around the world and are being utilized as platforms for various pilot studies and everyday use \citep{tsai2017learning,vsvabensky2026fifteen}.

It has been 15 years since then. Reflecting the passage of time and the evolving nature of LA practice and research, the definition of LA itself has been updated by \citet{SoLARDefinitionTaskforce2025}:
\begin{quote}
    Learning analytics is the collection, analysis, interpretation and communication of data about learners and their learning that provides theoretically relevant and actionable insights to enhance learning and teaching.
\end{quote}
Such revisions to the definition can be seen as a manifestation of LA's ongoing efforts to reexamine its own definition and sincerely confront the questions of what it is, what it focuses on, and what its purpose is, thereby reflecting these insights in its work.

However, these endeavors lack a critical consideration essential for the further development of the field: namely, its philosophical, and particularly metaphysical, foundations. LA certainly deals with ``data''; but what exactly is ``data''? LA is certainly ``the collection, analysis, interpretation and communication of data''; but who is doing what to whom, and how? LA certainly has ``to enhance learning and teaching'' as its ultimate goal; but can this truly be called the means of LA itself? For any mature field, the process of earnestly confronting its own ontology has been a rite of passage for its development and maturation. For example, economics has engaged in repeated philosophical or metaphysical debates (e.g., \citet{hausman1994philosophy,smith2005aristotle}), and in more recent fields, philosophy regarding generative AI and LLMs is thriving \citep{morgan2024dragon,gourlay2024generative}.

Some might argue that such detached philosophical reflections are unnecessary for this endeavor, which is driven by the goal of enhancing learning and teaching. However, this would hinder the establishment of LA's own academic and practical foundations, and would leave the various activities built upon them lacking in identity and explanatory power. Furthermore, as we saw earlier, the definition of LA itself has been either a top-down, organizational one provided by SoLAR or a bottom-up one derived from literature reviews (e.g., \citet{khalil2022comparison,gavsevic2015let}). Even though it is a field characterized by \textit{bricolage} \citep{gavsevic2017piecing,joksimovic2019journey}, the endeavor to construct LA's own identity based on its own principles is an indispensable task for providing a foundation for academic and practical activities.

This paper is an attempt at metaphysical reflection on LA. In this process, this paper argues that LA has, in significant portions of its practice, systematically conflated its purpose with its operations. This conflation is not merely a conceptual imprecision; it represents an ontological misalignment with what LA's own definition commits it to. Rendering this structure visible is the primary task of descriptive metaphysics \cite{Strawson1959Individuals} as applied here; rather than imposing our own system from the outside, it reveals the intrinsic nature of LA research and practice from within. As a result, this paper demonstrates both a metaphysical system that encompasses the history of LA research and practice, and the fact that certain practices within LA stand in a state of tension with its metaphysical conditions.

\section{Previous Work}
\subsection{Philosophical Reflections on LA}
Philosophy has often been discussed in terms of three broad domains: Axiology (the study of value), Epistemology (the study of knowledge), and Metaphysics, which studies beyond the physical world \citep{dennis2020branches,aliyu2015ontology}. Since LA is a field concerned with learners' learning, the inclusion of epistemology was essential. Furthermore, since LA is a field that involves practical application in educational settings, there is a critical need for discussion regarding its ethical and social implications; both of these areas have been the subject of vigorous debate.

First, examples of LA addressing epistemology include \citet{knight2013epistemology} and \citet{knight2014epistemology}. These works view epistemology as a driving force behind assessment---which uncovers learners' knowledge---and pedagogy---which builds high-quality knowledge---and discuss the triangular relationship among these three elements. Furthermore, \citet{doroudi2024paradigms} discusses the LA paradigm from the perspective of the relationship between machine learning and epistemology. He argues for shifting the focus from the correctness to the usefulness of constructed machine learning models and supports the importance of pursuing model robustness.

Discussion on the philosophy of LA are particularly active in the realm of ethics. This is closely linked to the fact that LA has a history of developing in tandem with educational practice. \citet{slade2013learning} constructed an ethical framework that views LA as part of the moral practice of education, emphasizing that students are active collaborators in the educational process and that educational institutions bear a responsibility to use data to improve learning outcomes. \citet{greller2012translating} position ethics as one of the key ``external constraints'' necessary to ensure the successful and beneficial use of educational data. \citet{kitto2019practical} emphasize the practicality of ethics in LA and propose an approach that addresses challenges through rational action grounded in the principles of virtue ethics.

In connection with these ethical debates, LA itself has a history of developing under the teleological premise of improving learning and education. As a result, much LA research and practice has been designed to incorporate normative judgments regarding educational decisions and the processes for making them within the LA framework. The ``learning analytics cycle'' proposed by \citet{clow2012learning}, consisting of learners, data, metrics, and interventions, explicitly embeds within itself the process of providing feedback on analysis results to learners and taking actions that influence them. \citet{khalil2015learning} explicitly reference this, incorporating an ``Act'' phase into their LA lifecycle that encourages educational decision-making based on insights gained through LA. \citet{mor2015learning} conceptualize a ``virtuous circle'' in which data generated by LA feeds back into the teacher's inquiry, leading to improvements in learning design. Furthermore, the concept of ``prescriptive learning analytics''---which actively asks what LA can do for learners to achieve desirable learning and educational outcomes---has been proposed \citep{susnjak2022learning,susnjak2024beyond}. Some research uses the term ``evidence-based'' to describe the nature of insights derived from LA and supports proactive interventions for learners based on such ``evidence-based'' insights (e.g., \citet{rienties2016analytics4action, kim2016toward, yan2024evidence}). In fact, this approach is supported by an ethical stance that argues that, since data processing by LA already involves value judgments, LA should be designed as a normative practice with an awareness of its normative nature (cf. \citet{slade2013learning}).

However, despite this accumulation of research and practice, there has been very little discussion regarding the metaphysics of LA, particularly its ontology. In other words, it cannot be said that metaphysical considerations regarding the ontological nature of the data that LA operates on, or the legitimate grounds for inference, have been explicitly addressed to date. This relative absence of metaphysical inquiry may stem from LA's historical development as a practice-oriented and methodologically driven field, in which the operational usefulness of data was often prioritized over reflection on the ontological status of data itself. Providing answers to such metaphysical questions would make a significant contribution to establishing the foundation and forum for open discussions regarding the epistemology and axiology of LA.

\subsection{Data Concepts in LA}
In this paper, in order to examine the metaphysics of LA, we will also discuss the data that LA manipulates and uses as the basis for its inferences. Therefore, providing an overview of how data has been treated in LA will be useful in offering insights for the subsequent discussion.

\citet{ballsun2012asking} argued that data is not a concept with a single, fixed definition, but rather a ``socially constructed concept'' that varies depending on an individual's background, role, and context. In contrast, it was \citet{floridi2013philosophy} who defined data from the perspective of \textit{diaphora} (meaning ``difference'' in Greek). He defined a datum as ``a putative fact regarding some difference or lack of uniformity within some context.'' Furthermore, in his General Definition of Information, he states that information consists of meaningful data organized according to logical forms. Based on the definitions provided by these two scholars, it is suggested that the nature of data as knowledge involves some form of structure and formalization.

The fact that some forms of LA, as discussed in the previous section, advocate for an ``evidence-based'' approach can be interpreted as reflecting the belief that data can serve as objective evidence and possesses a value that human intuition or observation cannot provide. At the same time, it is also true that skepticism is being directed toward the view that data reflects an objective and neutral truth. In fact, the treatment of data in LA to date often includes criticism and reflection on the assumption that data is neutral and objective. There are frequent calls to reconsider the way LA has often treated data as neutral, as well as to reexamine the power relations and normativity inherent in data, and the political nature of data collection, analysis, and application \citep{selwyn2020re,prinsloo2022answer}. Furthermore, there are epistemological suggestions that data should be interpreted based on theory within causal, social, and technical contexts, rather than being treated as objective facts \citep{rogers2015critical}.

Such critical analyses of data handling in LA clearly reflect the multifaceted nature of data. Nevertheless, it is difficult to say that the LA community has engaged sufficiently in discussions questioning the ontological nature of data. Examining the ontology of the data that LA directly manipulates would ultimately provide a metaphysical framework for understanding the metaphysical meaning of LA---that is, ``what LA is actually doing.''

\section{Metaphysical System for LA}
\subsection{Descriptive Metaphysics as a Methodology}
\begin{figure}[htbp]
    \centering 
    \includegraphics[width=0.9\linewidth]{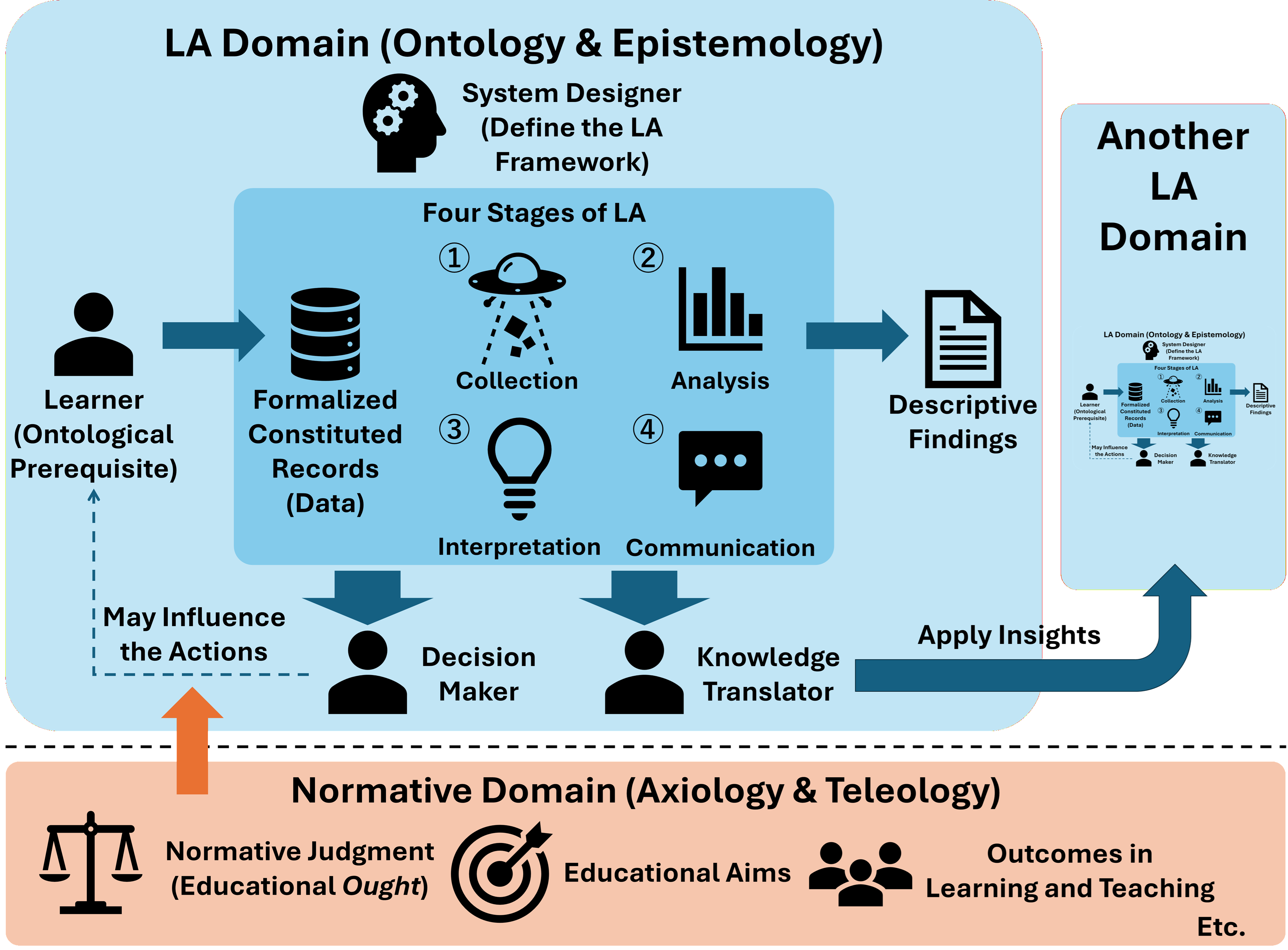}
    \caption{Summary of the metaphysical structure of LA}
    \label{fig:meta-la}
\end{figure}

In this section, we attempt to establish a metaphysical system for LA. Specifically, we attempt to provide an answer to the metaphysical question of what kind of endeavor LA is, using the framework of descriptive metaphysics by \citet{Strawson1959Individuals}. Descriptive metaphysics is a concept that stands in contrast to revisionary metaphysics. Whereas revisionary metaphysics seeks to challenge existing common sense and frameworks in order to provide a better explanation of existence, descriptive metaphysics attempts to analyze and elucidate the foundations of the common sense and conceptual structures we actually possess. The role of this section is not to impose a better ideal form on LA, but rather to simply describe the foundations of the conceptual structure inherent in LA. The metaphysical structure of LA proposed in this section is summarized in Figure \ref{fig:meta-la}.

As \citet{Strawson1959Individuals} conceived it, descriptive metaphysics does not merely inventory existing practices; it reveals the conceptual commitments that those practices presuppose including commitments that practitioners themselves may not have made explicit or that existing practices may have violated. The identification of tension between norm-embedded LA (detailed in Section \ref{sec:norm-embedded}) and the first principle is therefore not a revisionary imposition but a descriptive finding: it surfaces a structural inconsistency already latent in LA's own self-understanding.

\subsection{Refining the Concept of ``Data''}\label{sec:concept-of-data}
\subsubsection{The Non-Triviality of ``Data'' in LA}
As stated in the definition given earlier, LA is ``the collection, analysis, interpretation and communication of data about learners and their learning that provides theoretically relevant and actionable insights to enhance learning and teaching.'' It is true that the definition of LA can change with the times. However, rather than viewing this definition as absolute, it is more appropriate to regard it as a shared understanding of LA that has been built up within the community over time: that is, this paper treats the SoLAR definition as a privileged articulation of LA's self-understanding. Since this definition will be referred to frequently throughout this paper, let us formalize it as a single definition.

\begin{dfn}\label{dfn:what-is-la}
    LA is ``the collection, analysis, interpretation and communication of data about learners and their learning that provides theoretically relevant and actionable insights to enhance learning and teaching.''
\end{dfn}

First, let us carefully consider the concept of ``data'' that LA operates upon. What ontological character does the ``data'' appearing in Definition \ref{dfn:what-is-la} possess? In other words, what kind of entity is ``data''?

\citet{hanson1979patterns} argued that we do not view facts objectively as they are, but rather interpret reality through the lenses of our own assumptions, theories, and knowledge: a phenomenon he termed the ``theory-ladenness of observation.'' This raises a question for LA: is there a theory, or the absence thereof, that serves as the basis for deciding which information a given LA system should or should not record as data? For example, consider an LA practice that records login activities and their timestamps. Why are those specific activities selected, while the learner's pulse of learning environment are not? This is because there exists a theoretical judgment behind that LA practice that justifies such a selection. However, this question concerns the epistemological origin of data, ``what we can know about learning from data,'' and thus operates on a different level from the metaphysical question addressed in this paper. Therefore, this paper actively chooses methodological neutrality, holding that the data handled by LA can be treated as an object of study regardless of whether it is theory-laden, and positions theory-ladenness questions outside the metaphysics of LA.

Next, is the ``data'' in LA something given, or does it derive from some other entity? In definition \ref{dfn:what-is-la}, the term ``data'' is immediately modified by the phrase ``about learners and their learning.'' This strongly implies that ``data'' derives from ``learners and their learning.'' To begin with, the phrase ``data about learners and their learning'' in the definition does not simply mean that the data provides a description of learners and their learning. This is because if learners as entities and the phenomenon of learning did not actually exist, no data pertaining to them would exist. Therefore, data is a record ``constructed'' from learners and their learning; it does not merely serve the role of describing or explaining learners and their learning. Furthermore, the subject of the act of learning is clearly the learner, and the phenomenon of learning cannot exist without learners. Consequently, the existence of learners is the ontological precondition for ``data'' in LA, and thus the ontological precondition for LA itself. Based on the above considerations, it is appropriate to introduce the following ``zeroth principle'' into the metaphysics of LA.

\begin{prin}[The zeroth principle]\label{prin:the-0th}
    The learner is an ontological prerequisite for LA.
\end{prin}
This principle is not an external justification derived from LA's objective of enhancing learning and teaching, but rather an intrinsic justification derived directly from the manipulation of data---the essential operation of LA. In other words, the reason why learners are an ontological prerequisite is not because ``LA aims to improve learning and teaching,'' but because LA is, in and of itself, the act of manipulating data. This implies that it does not rely on a normative or moral proposition such as ``the learner should be at the center of learning improvement.''

Finally, we discuss the metaphysical form that ``data'' should take within LA. In Definition \ref{dfn:what-is-la}, four external operations on ``data''---``the collection, analysis, interpretation and communication''---are defined. Therefore, ``data'' in LA is required to be capable of undergoing these four types of operations. In practice, this is formalized by technical data specifications such as Experience API (xAPI) \citep{xAPIcom_2026} and Caliper \citep{IMSGlobal_Caliper_2026}. However, these technical specifications should not be understood as constituting the ontological nature of data within the metaphysics of LA; rather, they should be understood as constraints imposed a posteriori to satisfy the requirements imposed on the nature of data.

\subsubsection{Data as Formalized Constituted Records}
Based on the above analysis, ``data'' in LA is precisely defined as follows:
\begin{dfn}[Refining data concepts]\label{dfn:data}
    The ``data'' in LA is formalized constituted records.
\end{dfn}

The term ``constituted'' here indicates that data does not rely on the ``Myth of the Given'' described by \citet{sellars1963science}. In other words, it means that data is not a purely given entity, but rather a construct produced from the ontological precondition of the learner. Specifically, the data is constrained by a conceptual framework in the sense that it reflects entities from which some form of abstraction has been performed regarding the learner and their learning process, and it is constrained by a technical framework in the sense that it is subject to the condition of whether or not it can be captured by an LA system. The term ``formalized'' indicates that the constituted record undergoes a process of structuring that enables it to undergo the ``collection, analysis, interpretation and communication.'' In other words, ``constitution'' precedes ``formalization,'' not the other way around. These two are ontologically distinguishable operations.

Data in LA arises through two reductions: constitution, which abstracts learner events through conceptual and observational selection, and formalization, which structures those records into operable forms. LA data is therefore separated from learning events by these two transformations.

\subsection{The First Principle: The Minimal Legitimate Ground of Inference}\label{sec:the-1st}
Based on Definition \ref{dfn:what-is-la}, Principle \ref{prin:the-0th} (the zeroth principle), which specifies the ontological preconditions in LA, and the refined definition of data (Definition \ref{dfn:data}), we formulate the first principle of LA's metaphysics as follows.

\begin{prin}[The first principle]\label{prin:the-1st}
    The minimal legitimate ground of inference in LA is the formalized constituted records about learners and their learning.
\end{prin}

This first principle can be derived from the definition of LA and the zeroth principle. Since the learner is an ontological prerequisite for LA, the basis for LA's inference and its origin are limited to the learner. Taking this into account, along with Definition \ref{dfn:what-is-la}, it follows that the basis for LA's inference is data concerning the learners and their learning. Furthermore, Definition \ref{dfn:what-is-la} does not entail the manipulation of anything other than data about the learners and their learning. Accepting any non-recorded basis of inference would require introducing entities not licensed by the definition itself. Therefore, LA can be formulated as the endeavor to perform operations on formalized constituted records about the learners and their learning---using only such records as legitimate grounds---in order to obtain ``theoretically relevant and actionable insights.''

Furthermore, the first principle serves to strictly define LA as an epistemological practice. In other words, inferences regarding unrecorded learning processes fall outside the scope of LA. Consequently, normative judgments, value assessments, and ethical judgments, which are not ontologically embedded in the records, are not incorporated as grounds for inference within LA. From the perspective of ethics in LA, or from the stance that LA should be utilized in educational decision-making to create a feedback loop for improving learning and education in the classroom, this may seem unjustified as it narrows the scope of LA. However, this actually functions as a safeguard against the is/ought problem proposed by Hume \citep{hume1739treatise}. That is to say, a record that describes the phenomenon of learning---even while reducing it---cannot, through any logical inference, transcend its role as an ``is'' proposition describing the phenomenon. Consequently, it is logically unfeasible to derive normative lessons or guidance for learners, or to formulate ``ought'' propositions such as ``what constitutes desirable learning'' or ``what constitutes successful learning,'' from the facts obtained from the record. When making educational judgments, one must inevitably make some form of value judgment; this means that we must always supplement fact-based information with our views on what is considered desirable \citep{biesta2015good}. We must discuss such normative, value-based, and ethical judgments not within LA itself, but outside of it.

It is also important to note that the first principle does not itself rest on a normative proposition; it does not claim that LA \textit{ought} to respect the is/ought distinction. Rather, it observes that the definition of LA, as an operation on formalized constituted records, provides no logical mechanism by which normative propositions could enter as grounds of inference. This is an analytic observation about the logical structure of the definition, not a normative injunction.

\subsection{The Ontological Character of ``Formalized Constituted Records''}\label{sec:fcr}
Thus far, we have derived the ontological status of data in LA, as well as the zeroth and first principles of LA, from the definition of LA. In this section, building on the discussion so far, we will discuss the ontological nature of data as ``formalized constituted records.''

An artifact qualifies as a formalized constituted record if and only if it satisfies two conditions: (1) it is constituted from learner-generated events through conceptual and observational reduction, and (2) it is structured to be operable by an LA system; that is, it can serve as the subject of collection, analysis, interpretation, or communication as defined in Definition \ref{dfn:what-is-la}. Teacher observation notes or learner self-reports become formalized constituted records when, and only when, they are subjected to these operations within as LA system. Prior to such formalization, they exist as pre-data entities outside the scope of LA's metaphysics.

Such characteristics reveal the following facts about the data in LA. First, data in LA is not something given; rather, it is always produced by the learner. This is a natural consequence of recognizing only the learner as the ontological precondition of LA. As mentioned earlier, the data handled in LA always pertains to ``learners and their learning,'' and is further reduced in two ways: constitutively and formally. Consequently, data in LA is never treated as ``learners and their learning'' in and of itself.

This provides insight into the meaning of data in LA: the meaning of the data is not determined without reference to the conceptual and formal frameworks that constitute it. For example, in a system such as that described by \citet{boticki2019book}, information such as reading time, the number of events, the use of markers and notes, and completion rates becomes the subject of analysis and interpretation, while other information such as the learner's eye movements and the learning environment is disregarded. However, in systems such as those described by \citet{vatrapu2013eye}, on the other hand, the learner's eye movements become the subject of analysis, and even emotional or cognitive characteristics are interpreted from them, while other information is disregarded. Whereas the meaning of the data in the former case relates to the learner's interaction with the system, the data in the latter case relates to the learner's unconscious actions. Thus, the meaning of the data is determined only by reference to the context and framework of LA. In other words, there is no one-to-one correspondence between the obtained data and learning events.

\subsection{Agents and Their Ontological Status}\label{sec:agents}
\subsubsection{The Identification of the Agents}
Now that we have formalized the principles of LA, we will move on to identifying the agents involved in LA. To summarize initially, based on the definition in \ref{dfn:what-is-la} and its purpose (``to enhance learning and teaching''), we can identify eight agents involved in LA, as shown in Table \ref{tbl:agents}. The taxonomy is generated by decomposing the operations specified in Definition \ref{dfn:what-is-la} and the teleological structure attached to them.

\begin{table}[htbp]
    \centering
    \caption{Summary of the Agents in LA}
    \label{tbl:agents}
    \begin{tabular}{lp{230pt}}
        \toprule
        Agent & Essential Function \\ 
        \midrule
        Learner & The subject of data and its producer. An ontological precondition for the LA process. May also serve as the Decision Maker. \\
        System Designer & The designer of the LA system. A structural precondition for the LA process, defining the framework of data relationships. The highest level of responsibility attribution. \\
        Data Collector & Producer of constitutive records. While the System Designer defines the framework for collection, the Data Collector makes collection decisions within that framework. \\
        Data Analyzer & Performer of analysis. Bears a duty-based responsibility for selecting analytical methods. \\
        Data Interpreter & The executor of interpretation. Like the analyst, bears a duty-based responsibility for selecting interpretive methods. \\
        Data Communicator & The communicator (sender) of findings. The selection of communication format, audience, and timing influences the reception of findings. \\
        Decision Maker & The agent that directly incorporates LA findings into their own practical judgments.  \\
        Knowledge Translator & The entity that processes, develops, and translates findings into different contexts. Calculates inputs for other LA processes. \\
        \bottomrule
    \end{tabular}
\end{table}

First, it follows directly from the zeroth principle that learners exist as independent agents. This is the minimal ontological prerequisite in LA, and learners are the only agents capable of generating data as ``formalized constituted records.'' In this paper, we treat learners solely as agents that are the subjects of records and produce records; however, as will be discussed later, this does not imply that we deny learners' agency or decision-making authority.

Next, it is necessary to establish the system designer, who determines the framework within which a given LA system is designed, as a distinct agent. The system designer is, literally, the agent responsible for the design and development of the LA system and, within a specific LA practice, determines the framework for data manipulation and data reduction. In other words, the system designer is a prerequisite for the LA system. The reason this entity is established as a distinct agent is that it must be defined as the highest-level entity to which responsibility for events occurring within a given LA system is attributed. Although system designers are not included in the data flow, they are included as the ultimate party responsible within the framework of responsibility and causality that defines the entire LA system.

After establishing an agent that defines the entire LA system, since the definition of LA (Definition \ref{dfn:what-is-la}) states that LA is ``the collection, analysis, interpretation and communication of data,'' it is ontologically necessary to establish separate entities responsible for each of these four operations. While in practice, researchers often handle all four of these tasks, it is beneficial for future discussions to consider them as distinct agents from an ontological perspective.
\begin{itemize}
    \item Data collector: The agent responsible for data collection. The framework for data collection carried out by this entity is defined by system designers. In this context, they bear responsibility for the necessity and sufficiency of the data collected.
    \item Data analyzer: The agent that performs analysis on the data collected by the data collector. They possess the ability to judge the selection and validity of analytical methods and bear a duty of responsibility for them.
    \item Data interpreter: The agent that interprets the data analyzed by the data analyzer to derive descriptive insights. Like the data analyzer, they possess the ability to judge the selection and validity of interpretation methods, particularly regarding whether the insights obtained are ``theoretically relevant and actionable'', and bear a duty of responsibility for these matters.
    \item Data communicator: The agent that transmits the interpretations made by the data interpreter to another agent. They possess the ability to make methodological choices regarding the form of communication (such as format, recipient, and timing) as well as the accuracy and integrity of the content being communicated, and bear responsibility for these decisions.
\end{itemize}
As discussed in Section \ref{sec:the-1st}, it is logically unfeasible to derive normative (i.e., ought-based) propositions regarding desirable learning or learners from descriptive (i.e., is-based) propositions derived from data. Therefore, these four agents also bear the responsibility of ensuring that normative, value-based, and ethical judgments, which would render the LA logically ``impure,'' do not become mixed into its internal operations.

Decision makers are also one of the key agents in LA. In LA, in order to achieve the goal of ``enhancing learning and teaching,'' the insights gained must be reflected in the voluntary decisions and practical judgments of other stakeholders. In this context, decision makers are permitted to bring about changes in learning and teaching based on their own independent judgments, while taking into account the insights provided by data communicators. Specifically, learners, instructors, and policymakers fall into this category. Learners make independent judgments to enhance their own learning or that of other learners. Instructors, as entities capable of influencing learners' decision-making, make independent judgments regarding teaching and guidance. Policymakers, while not directly involved in the learning or teaching processes, make independent judgments to improve the frameworks for learning and teaching.

Finally, it is necessary to include a ``knowledge translator'' as an agent who processes and develops insights obtained from data communicators, or translates them into different contexts. While the seven agents defined so far are all agents that make decisions within a single LA system, the knowledge translator is the only agent capable of providing insights to a different LA system. As mentioned in Section \ref{sec:fcr}, the meaning of data in LA is not determined without referring to the framework that constitutes it. In other words, the role of the knowledge translator is essential to transform insights obtained within one LA system into a form that can be utilized by another LA system with a different framework. Specifically, the knowledge translator makes the insights and their meanings derived from the data accessible for reference even outside the LA system to which they belong. In practice, researchers who publish their findings externally in the form of research papers are prime examples of this role.

\subsubsection{Defining the Boundaries of the Agents}
In this framework, the concept of the ``teacher'' has not been included as an independent agent. Many readers may have questions or objections regarding this. However, this approach can be deemed valid based on the following three grounds. First, the act of learning can take place even without the presence of a teacher. Learning can proceed even in the absence of a teacher, as long as the learner exercises their agency by reading on their own or using the system. Therefore, the teacher is not an ontological prerequisite for LA and does not possess the subjective status derived from the zeroth principle that the learner possesses. Consequently, in this paper, the teacher is subsumed under the category of decision makers: specifically, those capable of influencing the learner's autonomous choices, thereby enabling the description of the teacher's impact on learning within the metaphysical framework of LA.

Second, the exclusion of the teacher as an independent agent within LA reflects a structural claim about LA's operational definition: since Definition \ref{dfn:what-is-la} specifies ``data about learners and their learning'' as LA's object of operation, data generated solely by teachers, which is independent of learner activity, falls outside LA's operational scope by definition. The teacher's role, particularly the exercise of pedagogical judgment, belongs structurally to the normative domain that this framework locates outside LA. Positioning the teacher's judgment there is not a demotion but a protection: it preserves the teacher's normative authority from being subsumed into LA's descriptive operations.

Third, the definition of LA (Definition \ref{dfn:what-is-la}) states its purpose as ``to enhance learning and teaching,'' which clearly includes ``teaching.'' However, in the same definition, the object of LA's operations is specified as ``data about learners and their learning,'' and the terms ``teacher'' or ``teaching'' do not appear there. If interpreted literally, this implies that teachers are not positioned as a source of data within the definition of LA. Thus, the definition of LA inherently contains a tension between teleology and ontology. However, this tension is not an internal contradiction; rather, it reflects the fact that LA, as an epistemological practice, relies on data produced by learners and their learning, and does not deal with data derived solely from the teacher. Identifying the teacher as an independent agent within LA leads to an unreasonable obscuring of this tension.


\subsection{The Critical Function of Systems}
The metaphysical system developed in Sections \ref{sec:concept-of-data}–\ref{sec:agents} performs two critical functions with respect to LA practice: it delineates the legitimate basis for inference, and it makes normative contamination detectable at the agent level.

\subsubsection{Limits on Inference}
As we saw in Sections \ref{sec:concept-of-data}-\ref{sec:fcr}, the sole legitimate basis for inference in LA is the ``formalized constituted records.'' This is an entity of a descriptive nature that reduces learners and events related to their learning in two stages. Therefore, normative judgments, value assessments, and ethical judgments are not derived from it. In other words, these do not enter into the LA as grounds for inference. This is a proposition that serves as the ontological elaboration of the first principle of the LA.

To give a concrete example, suppose a group of learners used the LA system for 10 hours a week and their academic performance improved. In the metaphysics of LA proposed in this paper, even if a descriptive inference such as ``using the system for 10 hours a week improves performance'' is possible from such descriptive facts, it does not derive the normative inference that ``therefore, learners should use the LA system for 10 hours a week.'' Similarly, in the identification or detection of ``at-risk students'' \citep{akccapinar2019using,foster2020effectiveness}---a common practice in LA---even if a descriptive inference such as ``students who meet certain conditions exhibit tendencies toward being at risk'' can be derived, a normative inference such as ``therefore, students should study to avoid falling into such tendencies'' is impossible. In order to make such normative inferences, LA must borrow normative propositions from sources outside its own metaphysics—such as educational philosophy or educational ethics.

\subsubsection{Detectability of Normative Contamination}
Furthermore, by identifying eight types of agents involved in LA, it becomes possible to detect at which point in the LA process the normative bias described above was introduced. This detectability of normative contamination is not merely an analytical property of the system; it has substantive implications for how we evaluate existing LA practices. These implications are taken up in Section 4.

\section{Discussion}
\subsection{The Boundary of LA and the Locus of Normative Judgment}\label{sec:norm-embedded}
In this paper, LA posits as its first principle that the sole legitimate basis for its reasoning is limited to data: that is, ``formalized constituted records.'' This limitation does not unduly narrow the scope of LA; rather, it explicitly articulates the ontological conditions necessary for LA to function as an epistemological practice. In this section, we discuss the positive significance of this demarcation.

The definition of LA (Definition \ref{dfn:what-is-la}) states its purpose as ``to enhance learning and teaching.'' From this teleological definition, it is sometimes argued that ``LA should contribute directly to the improvement of learning and teaching, and to that end, it should incorporate normative judgments regarding what constitutes desirable learning and what kinds of interventions are effective.'' Based on this argument, some LA systems are designed to incorporate a definition of desirable learning states into its system design in advance, and evaluates learners and intervenes in learning and teaching in light of that definition. That is, it ``embeds'' norms within itself. In this context, LA systems that use the term ``evidence-based'' to reflect the idea of deriving norms from facts (e.g., \citet{kim2016toward, rienties2016analytics4action, yan2024evidence}) can be examples of such LA.

Furthermore, while LAs that form feedback loops (e.g., \citet{clow2012learning, khalil2015learning}) and prescriptive LAs (e.g., \citet{susnjak2022learning, susnjak2024beyond}), while structurally value-neutral, include behavioral change aimed at optimizing learning and teaching within their frameworks; thus, depending on the designer's intent, definitions of desirable learning states may be incorporated into the system design prior to its development. Consequently, it can be said that these frameworks tend to adopt structures that are more prone to embed norms within itself than unidirectional LA.

Here, we define ``norm-embedded LA'' as LA in which normative propositions about desirable learning states are incorporated as \textit{prior constraints on data collection, analysis, or interpretation}, rather than as inputs to decision making processes that operate outside the LA system. We argue that the norm-embedded LA poses an ontological challenge to the first principles of LA by conflating the purpose of LA with its operations. The fact that LA aims to improve learning and teaching as its ``purpose'' and the fact that it places the ``means'' for achieving that purpose within itself are ontologically distinct issues. The realization of this purpose depends on the normative judgments made by the decision maker---who receives the descriptive findings calculated by LA---outside the scope of LA. Furthermore, by embedding normative judgments and the processes involved in making them, norm-embedded LA incorporates a definition of ``desirable learning behavior'' into the system design, evaluates learners and instructors against that definition, and intervenes in practice. In this case, normative judgments existentially precede the process of constructing the data on which the LA bases its reasoning, resulting in an ontological tension with the first principle proposed here. The metaphysics of the LA describes the division of labor between the production of descriptive findings and normative judgments; it does not deny the value of the LA's purpose.

Let us look at a few concrete examples of problems caused by norm-embedded LA. Consider a case where the judgment that ``students who meet certain conditions exhibit at-risk tendencies'' leads to the introduction of the judgment that ``students should avoid falling into such tendencies.'' If the data analyzer held the normative belief that ``being at-risk is undesirable,'' s/he might introduce that belief into the data analysis, resulting in a biased analysis that focuses only on variables useful for detecting at-risk students while dismissing everything else as noise. In the other case, if the data interpreter held the value-based belief that ``scalable and efficient support is crucial for reducing the number of at-risk students,'' s/he might focus solely on aspects of the analysis results that can be interpreted in terms of scalability and efficiency, thereby overlooking other features. Norm-embedded LA may distort the essential processes of ``collection, analysis, interpretation and communication'' of data, which is inherent in LA.

Nevertheless, from a perspective that is keenly focused on ensuring the practical utility of LA, there may be criticism of the fact that LA does not allow for normative judgments. Yet this criticism can be refuted as follows: the descriptive insights derived from the ``formalized constituted records'' produced by LA serve as an indispensable epistemological foundation for decision makers to make normative judgments. In other words, the practical utility of LA can be achieved by rigorously generating descriptive insights that maximize the quality of normative judgments made outside of LA, even without incorporating those judgments into LA itself. In short, LA plays a role akin to a map in learning and educational practice. This map of learning and educational judgment does not itself indicate the destination; however, learners cannot reach their destination without the map.

Nonetheless, the following objection might be raised: ``Since LA itself assumes the theory-ladenness of data, it already entails a certain kind of normative judgment; therefore, LA itself must be an educational and moral entity that is conscious of functioning as a norm.'' However, this objection can also be refuted as follows. First, this type of objection is itself a fallacy in that it directly derives the ought-statement ``therefore, LA ought to be normative'' from the is-statement ``LA contains norms.'' Moral practice theory does not resolve such is/ought problems. Furthermore, the claim that the normativity of LA derives from the goal-driven nature or normativity of education (cf. \citet{biesta2015good}) does not stem from LA itself but borrows propositions from educational philosophy situated outside of LA; thus, it does not serve as a basis for LA's internal principles. Furthermore, if the aforementioned argument is attempting to derive the proposition that ``LA is, as a matter of fact, educational and moral,'' then it merely falls into the tautology of ``LA is normative; therefore, LA is normative.'' Such a proposition risks becoming entirely displacive; it forecloses rather than opens the inquiry into how normativity enters LA and at which point.

This metaphysical demarcation, it should be emphasized, does not diminish LA's practical contribution. Rather, it clarifies the conditions under which that contribution is epistemologically sound. Normative judgments and ethical discussions concerning LA are not eliminated but relocated: outside the metaphysics of LA (the discussion of ``what LA is''), where they can be conducted with the independence they require. 

\subsection{The Priority of the Learner and Its Implications}
Similarly, we will discuss the positive implications of LA treating the learner as its sole ontological precondition. This positioning means that, based on the zeroth principle, the learner is accorded a status that precedes the LA process, all agents, and the first principle.

The fact that the learner is an ontological precondition derived from the zeroth principle means that no design or practice of LA can infringe upon the precedence of the learner's existence. In other words, regardless of how the LA system is designed or utilized, the data that LA operates on must consist of records derived from the learner. Therefore, the criticism that ``viewing the learner solely as the subject and producer of data amounts to dehumanizing and objectifying the learner'' misunderstands the meaning of the concept of ontological priority. Rather, the learner's ontological priority provides a critical standard for determining whether treating the learner as a data-producing machine satisfies the metaphysical conditions of LA, and it serves as the metaphysical basis for the learner's protected status.

This does not, however, deny that learners are agents capable of making decisions and exercising autonomous judgment (cf. \citet{zimmerman2002becoming, viberg2020self}). As mentioned earlier, while learners may often serve as decision makers within the LA process, this is a secondary role for them, not their essential one. However, this merely means that, in describing the structure of the LA process, the description of the learner's decision-making capacity belongs outside the scope of LA. Questions regarding the learner's autonomy and agency are issues that should be addressed within the teleological and pedagogical discussions of LA. By securing a forum for such discussions outside its own framework, this metaphysical system does not suppress debates on learner autonomy and agency; rather, it safeguards the independence of those discussions.

Furthermore, a similar logic applies to critiques of the teacher's absence as an independent agent within LA. In this paper, the teacher is included solely as a form of decision maker and is not regarded as an independent agent. However, this does not imply a disregard for the teacher's educational role, nor does it negate the teacher's normative judgments. While the definition of LA states its purpose as ``to enhance learning and teaching,'' as long as it defines ``data about learners and their learning'' as its object of operation, data originating solely from the teacher is not included in LA's scope. This reflects the tension between teleology and ontology inherent in the definition of LA. And it is precisely when we confront this tension head-on that it becomes clear that the teacher's normative judgment fulfills the most crucial function of a ``decision maker''---one capable of influencing the behavior of the learner as a subject. Securing a space for the teacher's professional judgment outside the scope of LA does not negate the teacher's essential role; on the contrary, it metaphysically guarantees and respects the autonomy of the teaching profession.

\subsection{Relationships with Adjacent Fields}
LA is essentially a \textit{bricolage} field and has complex relationships with various adjacent fields \citep{gavsevic2017piecing,joksimovic2019journey}. In this section, using Educational Data Mining (EDM), learner modeling, and learning design as examples, we provide a framework for situating the structure of these adjacent fields within the ontology of LA.

\subsubsection{Educational Data Mining}
The boundary between EDM and LA has been the subject of ongoing debate since their inception \citep{siemens2012learning}. Although these fields have actively interacted with one another, the distinction between them has been primarily teleological: EDM has been described as aiming to discover insights from data, while LA aims to understand and optimize learning and educational practices by utilizing those insights. This framework adds an ontological perspective to this debate. Within the four-stage operational structure of ``collection, analysis, interpretation and communication'' defined by LA, EDM is positioned as the methodology for the ``analysis'' and ``interpretation'' phase. In other words, it can be argued that EDM is a methodological practice subsumed within the metaphysical structure of LA, and that LA functions as the ontological framework that determines the context of that methodology. This description does not pit EDM and LA against each other; rather, it contributes to metaphysically clarifying their respective roles and promoting their interrelatedness.

\subsubsection{Learner Modeling}
Learner modeling \citep{dillenbourg1992framework} is a related field closely connected to this framework. Learner modeling has aimed to represent learners' knowledge states, cognitive states, and behavioral patterns in a computable form. Based on this framework, such representations of learners can be understood as a descriptive reduction of the phenomenon of learning. However, when a constructed learner model functions as the basis for feedback, intervention, or recommendations to the learner, and when the learner adopts judgments based on that model---or, at times, even the model itself (cf. open learner models \citep{bull2010open})---as a framework for self-understanding and acts in accordance with it, the model ceases to be merely a description of the learning phenomenon and instead fulfills a generative function that produces learning events. In other words, the learner model deviates from its descriptive nature as a ``formalized constituted record'' and can function as a normative structure that determines the learner's learning behavior. By formulating this transformation ontologically, this framework makes explicit the question of descriptive neutrality in the learner model and, as a result, highlights the tension with the first principles of LA.

\subsubsection{Learning Design}
Learning design \citep{koper2006current} is a field that deals with the normative practice of designing learning activities. From the perspective of this framework, the design insights produced by learning design belong to the realm of normative judgments that lie outside of LA. However, learning design and LA are not entirely unrelated. When learning design is carried out within the LA process, its insights can influence the design judgments of the system designer through the knowledge translator. In other words, there exists a pathway through which the insights of learning design are not confined to normative judgments outside of LA, but are translated into design guidelines for the LA system. This provides a metaphysical description of the potential for collaboration between learning design and LA.

\subsection{Limitations}
No metaphysical system can avoid making choices and omissions in its construction. The LA metaphysics presented in this paper is no exception. Disclosing the limitations this metaphysics has will encourage critical dialogue regarding this system.

\subsubsection{The Value Neutrality of the System Itself}
This system claims to be descriptive and locates normative judgments regarding LA outside of itself. However, it has not been able to provide a definitive answer to the question of whether the system itself is, in the first place, value-neutral. For example, it is impossible to completely rule out the possibility that the very choices---such as treating the learner as an ontological precondition, placing the system designer at the apex of the chain of responsibility, or situating normative judgments outside of LA---reflect certain value-based or political stances. As \citet{Strawson1959Individuals} acknowledged, descriptive metaphysics inevitably foregrounds certain structures rather than neutrally reproducing them. The value-laden character of this framework is an invitation for critique, not a disqualification of the enterprise.

\subsubsection{Criteria for Identifying the Subjects}
This paper has identified eight subjects involved in LA, but it does not provide a complete answer to the question of why these eight in particular. In other words, the process of deriving the subjects itself involves choices. For example, while the decision not to treat institutions, organizations, and policymakers involved in LA as independent subjects is grounded in the definition of LA, a metaphysical argument to justify that choice itself has not been sufficiently developed. Questions regarding the identification of LA's subjects require more refined arguments and remain a task for future research.

\subsubsection{The Bricolage Quality of LA}
From its inception, LA has developed as a \textit{bricolage} field, drawing together concepts, methodologies, and theories from diverse disciplines such as cognitive science, educational technology, statistics, computer science, and the philosophy of education \citep{gavsevic2017piecing, joksimovic2019journey}. This bricolage-like nature is both the source of LA's richness and a reflection of the difficulty of encompassing its entirety within a single metaphysical framework. While the framework presented in this paper offers a coherent structure based on a definition of LA, the extent to which this framework can explain LA practices across diverse contexts---such as school education, higher education, vocational training, and informal learning---remains a matter for future empirical and theoretical examination. This framework is not a definitive description but is presented as a line of inquiry to be refined through dialogue within the community.

\section{Conclusion and Future Work}
This paper has organized and visualized the metaphysics that LA has implicitly constructed throughout its history as an attempt at descriptive metaphysics. Specifically, starting from LA's own definition, we derived the zeroth principle, which posits the learner as an ontological precondition, and the first principle, which treats data regarding the learner and their learning, namely ``formalized constituted records,'' as the sole legitimate basis for inference: thereby visualizing its metaphysical foundation. Furthermore, based on this, we identified eight agents related to LA and their relationships, and presented two critical functions derived from them: the limitation of grounds for inference and the detectability of normative contamination.

This metaphysical system, by narrowing the scope of LA, is by no means intended to negate the existence of existing LA or to limit the future form of LA. It is intended to secure a space for independent discussion of epistemology, axiology, and ontology by separating the layers of these debates, which had not been explicitly addressed and had consequently become intermingled. In particular, this metaphysical system attempts to explicitly separate descriptive findings and normative judgments, which have tended to be intermingled in past LA practices, and to describe their division of labor metaphysically. This has contributed to building a bulwark against the is/ought problem: namely, that ``norms cannot be derived from facts,'' without compromising the practical utility of LA.

Otherwise, some readers may regard parts of this framework as stating what already appears obvious within LA practice. However, one contribution of descriptive metaphysics lies precisely in rendering such implicit assumptions explicit, systematic, and philosophically justifiable. This paper therefore contributes not only by proposing a framework, but by articulating the ontological structure that LA has long presupposed.

The core claim of this paper is straightforward: LA derives its epistemic authority from formalized constituted records, and any practice that embeds normative judgments prior to that derivation operates outside LA's own metaphysical conditions. Rendering this structure explicit is not an act of restriction but of clarification. The framework proposed here does not aspire to authority; it aspires to be useful as a starting point for the critical dialogue the LA community is already engaged in and as a stable reference point for that dialogue to build upon.


\bibliography{bibliography}

@String{Computing = "Computing" }

@String{Academic = "Academic Press" }

@String{Springer = "Springer-Verlag" }

@proceedings{LAK11,
  title     = {Proceedings of the 1st International Conference on Learning Analytics and Knowledge},
  editor    = {Phillip Long and George Siemens and Grainne Conole and Dragan Gas\v{e}vi\'{c}},
  year      = {2011},
  isbn      = {9781450309448},
  publisher = {Association for Computing Machinery},
  address   = {New York, NY, USA},
  doi       = {10.1145/2090116},
  url       = {https://doi.org/10.1145/2090116}
}

@article{anderson2008end,
  title={The end of theory: The data deluge makes the scientific method obsolete},
  author={Anderson, Chris},
  journal={Wired magazine},
  volume={16},
  number={7},
  pages={16--07},
  year={2008}
}

@article{siemens2013learning,
  title={Learning analytics: The emergence of a discipline},
  author={Siemens, George},
  journal={American behavioral scientist},
  volume={57},
  number={10},
  pages={1380--1400},
  year={2013},
  publisher={SAGE Publications Sage CA: Los Angeles, CA}
}

@misc{SoLARWhatIsLA2025,
  author       = {{Society for Learning Analytics Research}},
  title        = {What is Learning Analytics?},
  year         = {2025},
  howpublished = {\url{https://www.solaresearch.org/about/what-is-learning-analytics/}},
  note         = {Accessed: 2026-05-24}
}

@inproceedings{tsai2017learning,
  title={Learning analytics in higher education---challenges and policies: a review of eight learning analytics policies},
  author={Tsai, Yi-Shan and Gasevic, Dragan},
  booktitle={Proceedings of the seventh international learning analytics \& knowledge conference},
  pages={233--242},
  year={2017}
}

@inproceedings{vsvabensky2026fifteen,
  title={Fifteen years of learning analytics research: topics, trends, and challenges},
  author={{\v{S}}v{\'a}bensk{\`y}, Valdemar and Borchers, Conrad and Fortuna, Elvin and Cloude, Elizabeth B and Ga{\v{s}}evi{\'c}, Dragan},
  booktitle={Proceedings of the LAK26: 16th International Learning Analytics and Knowledge Conference},
  pages={697--708},
  year={2026}
}

@misc{SoLARDefinitionTaskforce2025,
  author       = {{Society for Learning Analytics Research}},
  title        = {SoLAR's Learning Analytics Definition Taskforce Releases Report},
  year         = {2025},
  month        = jun,
  day          = {18},
  howpublished = {\url{https://www.solaresearch.org/2025/06/solars-learning-analytics-definition-taskforce-releases-report/}},
  note         = {Accessed: 2026-05-24}
}

@book{hausman1994philosophy,
  title={The philosophy of economics: An anthology},
  author={Hausman, Daniel M},
  year={1994},
  publisher={Cambridge University Press}
}

@article{smith2005aristotle,
  title={Aristotle, Menger, Mises: An Essay in the Metaphysics of Economics},
  author={Smith, Barry},
  journal={Philosophers of capitalism: Menger, Mises, Rand, and beyond. Oxford: Lexington Books},
  pages={199--222},
  year={2005}
}

@phdthesis{morgan2024dragon,
  title={The Dragon, the Witch and the Juggernaut: Towards a Philosophy of Generative AI},
  author={Morgan, William R},
  year={2024},
  school={University of California, Berkeley}
}

@inproceedings{gourlay2024generative,
  title={Generative AIs, more-than-human authorship, and Husserl’s phenomenological ‘horizons’},
  author={Gourlay, Lesley},
  booktitle={Networked Learning Conference},
  volume={14},
  number={1},
  year={2024},
  organization={Networked Learning Conference}
}

@inproceedings{khalil2022comparison,
  title={A comparison of learning analytics frameworks: A systematic review},
  author={Khalil, Mohammad and Prinsloo, Paul and Slade, Sharon},
  booktitle={LAK22: 12th international learning analytics and knowledge conference},
  pages={152--163},
  year={2022}
}

@article{gavsevic2015let,
  title={Let’s not forget: Learning analytics are about learning},
  author={Ga{\v{s}}evi{\'c}, Dragan and Dawson, Shane and Siemens, George},
  journal={TechTrends},
  volume={59},
  number={1},
  pages={64--71},
  year={2015},
  publisher={Springer}
}

@article{gavsevic2017piecing,
  title={Piecing the learning analytics puzzle: A consolidated model of a field of research and practice},
  author={Ga{\v{s}}evi{\'c}, Dragan and Kovanovi{\'c}, Vitomir and Joksimovi{\'c}, Sre{\'c}ko},
  journal={Learning: Research and Practice},
  volume={3},
  number={1},
  pages={63--78},
  year={2017},
  publisher={Taylor \& Francis}
}

@article{joksimovic2019journey,
  title={The journey of learning analytics},
  author={Joksimovi{\'c}, Sre{\'c}ko and Kovanovi{\'c}, Vitomir and Dawson, Shane},
  journal={HERDSA Review of Higher Education},
  volume={6},
  pages={27--63},
  year={2019}
}

@book{Strawson1959Individuals,
  author    = {P. F. Strawson},
  title     = {Individuals: An Essay in Descriptive Metaphysics},
  publisher = {Methuen},
  address   = {London},
  year      = {1959}
}

@article{dennis2020branches,
  title={The Branches of Philosophy.”},
  author={Dennis, Otto},
  journal={Rudiments of Philosophy and Logic},
  pages={67--95},
  year={2020}
}

@inproceedings{aliyu2015ontology,
  title={Ontology, epistemology and axiology in quantitative and qualitative research: Elucidation of the research philophical misconception},
  author={Aliyu, Aliyu Ahmad and Singhry, Ibrahim Musa and Adamu, Haruna and AbuBakar, Mu'Awuya Muhammad},
  booktitle={Proceedings of the Academic Conference: Mediterranean Publications \& Research International on New Direction and Uncommon},
  volume={2},
  number={1},
  pages={1054--1068},
  year={2015}
}

@inproceedings{knight2013epistemology,
  title={Epistemology, pedagogy, assessment and learning analytics},
  author={Knight, Simon and Buckingham Shum, Simon and Littleton, Karen},
  booktitle={Proceedings of the third international conference on learning analytics and knowledge},
  pages={75--84},
  year={2013}
}

@article{knight2014epistemology,
  title={Epistemology, assessment, pedagogy: Where learning meets analytics in the middle space},
  author={Knight, Simon and Shum, Simon Buckingham and Littleton, Karen},
  journal={Journal of Learning Analytics},
  volume={1},
  number={2},
  pages={23--47},
  year={2014}
}

@article{doroudi2024paradigms,
  title={On the paradigms of learning analytics: Machine learning meets epistemology},
  author={Doroudi, Shayan},
  journal={Computers and education: artificial intelligence},
  volume={6},
  pages={100192},
  year={2024},
  publisher={Elsevier}
}

@article{slade2013learning,
  title={Learning analytics: Ethical issues and dilemmas},
  author={Slade, Sharon and Prinsloo, Paul},
  journal={American behavioral scientist},
  volume={57},
  number={10},
  pages={1510--1529},
  year={2013},
  publisher={SAGE Publications Sage CA: Los Angeles, CA}
}

@article{greller2012translating,
  title={Translating learning into numbers: A generic framework for learning analytics},
  author={Greller, Wolfgang and Drachsler, Hendrik},
  journal={Journal of Educational Technology \& Society},
  volume={15},
  number={3},
  pages={42--57},
  year={2012},
  publisher={JSTOR}
}

@article{kitto2019practical,
  title={Practical ethics for building learning analytics},
  author={Kitto, Kirsty and Knight, Simon},
  journal={British Journal of Educational Technology},
  volume={50},
  number={6},
  pages={2855--2870},
  year={2019},
  publisher={Wiley Online Library}
}

@inproceedings{clow2012learning,
  title={The learning analytics cycle: closing the loop effectively},
  author={Clow, Doug},
  booktitle={Proceedings of the 2nd international conference on learning analytics and knowledge},
  pages={134--138},
  year={2012}
}

@inproceedings{khalil2015learning,
  title={Learning analytics: principles and constraints},
  author={Khalil, Mohammad and Ebner, Martin},
  booktitle={EdMedia},
  pages={1789--1799},
  year={2015},
  organization={Association for the Advancement of Computing in Education (AACE)}
}

@misc{mor2015learning,
  title={Learning design, teacher inquiry into student learning and learning analytics: A call for action},
  author={Mor, Yishay and Ferguson, Rebecca and Wasson, Barbara},
  journal={British Journal of Educational Technology},
  volume={46},
  number={2},
  pages={221--229},
  year={2015},
  publisher={Wiley Online Library}
}

@article{susnjak2022learning,
  title={Learning analytics dashboard: a tool for providing actionable insights to learners},
  author={Susnjak, Teo and Ramaswami, Gomathy Suganya and Mathrani, Anuradha},
  journal={International Journal of Educational Technology in Higher Education},
  volume={19},
  number={1},
  pages={12},
  year={2022},
  publisher={Springer}
}

@article{susnjak2024beyond,
  title={Beyond predictive learning analytics modelling and onto explainable artificial intelligence with prescriptive analytics and ChatGPT},
  author={Susnjak, Teo},
  journal={International Journal of Artificial Intelligence in Education},
  volume={34},
  number={2},
  pages={452--482},
  year={2024},
  publisher={Springer}
}

@article{rienties2016analytics4action,
  title={Analytics4Action Evaluation Framework: A Review of Evidence-Based Learning Analytics Interventions at the Open University UK.},
  author={Rienties, Bart and Boroowa, Avinash and Cross, Simon and Kubiak, Chris and Mayles, Kevin and Murphy, Sam},
  journal={Journal of Interactive Media in Education},
  volume={2016},
  number={1},
  year={2016},
  publisher={ERIC}
}

@phdthesis{ballsun2012asking,
  title={Asking about data: exploring different realities of data via the social data flow network methodology},
  author={Ballsun-Stanton, Brian},
  year={2012},
  school={UNSW Sydney}
}

@book{floridi2013philosophy,
  title={The philosophy of information},
  author={Floridi, Luciano},
  year={2013},
  publisher={OUP Oxford}
}

@article{selwyn2020re,
  title={Re-imagining ‘learning analytics’… a case for starting again?},
  author={Selwyn, Neil},
  journal={The Internet and Higher Education},
  volume={46},
  pages={100745},
  year={2020},
  publisher={Elsevier}
}

@article{prinsloo2022answer,
  title={The answer is (not only) technological: Considering student data privacy in learning analytics},
  author={Prinsloo, Paul and Slade, Sharon and Khalil, Mohammad},
  journal={British Journal of Educational Technology},
  volume={53},
  number={4},
  pages={876--893},
  year={2022},
  publisher={Wiley Online Library}
}

@inproceedings{rogers2015critical,
  title={Critical realism and learning analytics research: epistemological implications of an ontological foundation},
  author={Rogers, Tim},
  booktitle={Proceedings of the fifth international conference on learning analytics and knowledge},
  pages={223--230},
  year={2015}
}

@article{yan2024evidence,
  title={Evidence-based multimodal learning analytics for feedback and reflection in collaborative learning},
  author={Yan, Lixiang and Echeverria, Vanessa and Jin, Yueqiao and Fernandez-Nieto, Gloria and Zhao, Linxuan and Li, Xinyu and Alfredo, Riordan and Swiecki, Zachari and Ga{\v{s}}evi{\'c}, Dragan and Martinez-Maldonado, Roberto},
  journal={British Journal of Educational Technology},
  volume={55},
  number={5},
  pages={1900--1925},
  year={2024},
  publisher={Wiley Online Library}
}

@article{kim2016toward,
  title={Toward evidence-based learning analytics: Using proxy variables to improve asynchronous online discussion environments},
  author={Kim, Dongho and Park, Yeonjeong and Yoon, Meehyun and Jo, Il-Hyun},
  journal={The Internet and Higher Education},
  volume={30},
  pages={30--43},
  year={2016},
  publisher={Elsevier}
}

@book{hanson1979patterns,
  title={Patterns of discovery: An inquiry into the conceptual foundations of science},
  author={Hanson, Norwood Russell},
  year={1979},
  publisher={Cup archive}
}

@book{sellars1963science,
  title={Science, perception and reality},
  author={Sellars, Wilfrid},
  year={1963},
  publisher={London: Routledge \& Kegan Paul}
}

@book{hume1739treatise,
  author       = {David Hume},
  title        = {A Treatise of Human Nature},
  volume       = {3},
  part          = {1},
  chapter       = {1},
  year         = {1739},
  publisher    = {John Noon},
  address      = {London},
  note         = {Book III, Part I, Section I},
}

@book{biesta2015good,
  title={Good education in an age of measurement: Ethics, politics, democracy},
  author={Biesta, Gert JJ},
  year={2015},
  publisher={Routledge}
}

@article{akccapinar2019using,
  title={Using learning analytics to develop early-warning system for at-risk students},
  author={Ak{\c{c}}ap{\i}nar, G{\"o}khan and Altun, Arif and A{\c{s}}kar, Petek},
  journal={International Journal of Educational Technology in Higher Education},
  volume={16},
  number={1},
  pages={40},
  year={2019},
  publisher={Springer}
}

@article{foster2020effectiveness,
  title={The effectiveness of learning analytics for identifying at-risk students in higher education},
  author={Foster, Ed and Siddle, Rebecca},
  journal={Assessment \& Evaluation in Higher Education},
  volume={45},
  number={6},
  pages={842--854},
  year={2020},
  publisher={Taylor \& Francis}
}

@article{zimmerman2002becoming,
  title={Becoming a self-regulated learner: An overview},
  author={Zimmerman, Barry J},
  journal={Theory into practice},
  volume={41},
  number={2},
  pages={64--70},
  year={2002},
  publisher={Taylor \& Francis}
}

@inproceedings{viberg2020self,
  title={Self-regulated learning and learning analytics in online learning environments: A review of empirical research},
  author={Viberg, Olga and Khalil, Mohammad and Baars, Martine},
  booktitle={Proceedings of the tenth international conference on learning analytics \& knowledge},
  pages={524--533},
  year={2020}
}

@inproceedings{siemens2012learning,
  title={Learning analytics and educational data mining: towards communication and collaboration},
  author={Siemens, George and Baker, Ryan SJ d},
  booktitle={Proceedings of the 2nd international conference on learning analytics and knowledge},
  pages={252--254},
  year={2012}
}

@incollection{bull2010open,
  title={Open learner models},
  author={Bull, Susan and Kay, Judy},
  booktitle={Advances in intelligent tutoring systems},
  pages={301--322},
  year={2010},
  publisher={Springer}
}

@article{dillenbourg1992framework,
  title={A framework for learner modelling},
  author={Dillenbourg, Pierre and Self, John},
  journal={Interactive learning environments},
  volume={2},
  number={2},
  pages={111--137},
  year={1992},
  publisher={Taylor \& Francis}
}

@article{koper2006current,
  title={Current research in learning design},
  author={Koper, Rob},
  journal={Journal of Educational Technology \& Society},
  volume={9},
  number={1},
  pages={13--22},
  year={2006},
  publisher={JSTOR}
}

@misc{xAPIcom_2026,
  author       = {{Rustici Software}},
  title        = {xAPI.com Homepage: What is xAPI (the Experience API)},
  year         = {2026},
  howpublished = {\url{https://xapi.com/}},
  note         = {Accessed: 2026-05-25}
}

@misc{IMSGlobal_Caliper_2026,
  author       = {{1EdTech Consortium}},
  title        = {Caliper Analytics},
  year         = {2026},
  howpublished = {\url{https://www.1edtech.org/standards/caliper}},
  note         = {Accessed: 2026-05-25}
}

@inproceedings{vatrapu2013eye,
  title={An eye-tracking study of notational, informational, and emotional aspects of learning analytics representations},
  author={Vatrapu, Ravi and Reimann, Peter and Bull, Susan and Johnson, Matthew},
  booktitle={Proceedings of the third international conference on learning analytics and knowledge},
  pages={125--134},
  year={2013}
}

@article{boticki2019book,
  title={E-book user modelling through learning analytics: the case of learner engagement and reading styles},
  author={Boticki, Ivica and Ak{\c{c}}ap{\i}nar, G{\"o}khan and Ogata, Hiroaki},
  journal={Interactive Learning Environments},
  volume={27},
  number={5-6},
  pages={754--765},
  year={2019},
  publisher={Taylor \& Francis}
}
\bibliographystyle{plainnat}

\end{document}